# Effect of relative in-plane twisting in graphene bilayer on sensing using surface plasmon resonance

**Amrit Kumar, Manjuladevi V and R. K. Gupta***

Department of Physics, Birla Institute of Technology and Science, Pilani (BITS Pilani), Rajasthan -333031

*E-mail: raj@pilani.bits-pilani.ac.in

**Abstract**

Surface plasmon resonance (SPR) is generally observed by excitation of surface plasmon polaritons on the metal (Au/Ag) surface. In order to utilize the SPR phenomenon for sensing application, the metal surface is functionalized with suitable ligands. Although such functionalization can enhance the specific adsorption capability of the sensor however due to large thickness of the ligands, the plasmonic field of the metal surface becomes less sensitive towards the adsorption of analytes. In the next generation SPR based sensor, graphene can be utilized not only as plasmonic material but also a suitable ligand for attracting analytes through π-π interaction. In this article, we present our theoretical simulation studies on the observation of SPR phenomenon using graphene monolayer (MLG), bilayer graphene (BLG) and in-plane twisted layers of BLG (T-BLG) as plasmonic materials deposited over Zinc-Selenide substrate. The Kretschmann configuration under wavelength interrogation setup was simulated and SPR wavelength for graphene systems/water interface was estimated. The bio-sensing simulation was performed and the sensing parameters viz. sensitivity, figure-of-merit (FOM) and plasmonic field for different graphene systems were obtained. Interestingly, the excellent sensing parameters were found in T-BLG system with relative in-plane twist angle near to magic angle viz. 1°. The enhancement is due to strong coupling between the layers twisted at the magic angle. This study demonstrates that the monolayer, bilayer and twisted bilayer graphene can be employed as a standalone layer system for not only generation of plasmonic fields but also enhanced sensing due to its intrinsic π−π interactions with bio-analytes.

## 1. Introduction

Surface plasmon resonance (SPR) is an optical phenomenon which is extremely popular in the field of bio-sensing and bio-analysis owing to its label free-detection of analytes at a very high sensitivity and resolution [1, 2]. The SPR is generally employed as a refractive index (RI) measurement technique which can be utilized to detect the adsorption of even a few analytes on the active area of the sensor [3]. The sensing application using SPR essentially utilizes several enhancement factors viz. suitable functionalization of the active area and optimization of optics. There are numerous reports in literature showing the specific sensing application using the SPR phenomenon with suitable functionalization of the active area [4]. The Kretschmann configuration using prism/metal/dielectric interfaces for generating SPR conditions is highly sensitive and practically viable. In this configuration, the surface plasmon polaritons (SPP) will be generated by the evanescent wave of electromagnetic (EM) wave incident on the metal/dielectric interface. A maximum energy transfer from the incident EM wave to the SPP wave takes place at the resonance which can be detected by observing the minimum intensity of the reflected wave at the given condition. At resonance, the plasmonic field is expected to be very high. This field interacts with the dielectric medium over the metal surface. The plasmonic field gets perturb due to dielectric change over the metal surface due to the adsorption of analytes. This can shift the resonance condition which can be measured using the SPR instrumentation. This is the fundamental of any sensing and bio-analysis using the SPR phenomenon. Recently, fiber optic based SPR [5–7] sensors are being studied as they support better polychromatic incidence mode application. Although, despite being cost effective and easy to use, they are less reliable towards complex biosensing applications. The practical limitations such as fragile nature, difficulties in suitable functionalization and usage of low resolution spectrometer makes it difficult for bio-sensing applications [8]. Fiber optic based sensor shows field attenuation which reduces the signal-to-noise ratio to a large extent. However, the prism based Krescthmann configuration offers a very high resolution, sensitivity, stability and reliability [9]. In a usual SPR based sensor, gold is most universally used as plasmonic material for the generation of SPP wave as it possesses several practical advantages like chemically inertness, easy functionalization, and good stability in aqueous medium. However, pure gold surface has lower affinity towards many analytes. To enhance the performance, several developments have been reported like use of bimetallic (Ag-Au) layer [10, 11], nanohybrids metal/dielectric structure Au with thin oxide film coating [12], amorphous carbon on surface plasmon active silver substrate [13] etc. In order to enhance the sensitivity towards specific analyte, the surface of the gold can be functionalized with help of a suitable ligands. The available gold based SPR system employed for bio-sensing offers the figure of merit (FOM) to be 11.78 in aqueous media (RI = 1.33) [14]. Further in wavelength interrogation mode, optical fiber based SPR devices offers sensitivity 2280 nm/RIU for gold and 3710 nm/RIU for silver as sensing layer [15]. Although, it can be enhanced up to ~5000 nm/RIU by adding few graphene layers over the silver metallic platform. In case of biosensors, sometime the thickness of the ligands is large enough to induce any perturbation in the plasmonic field due to adsorption of analytes. This reduces the sensing performance of the SPR based biosensors. Recently, few works have been reported [16–18] wherein graphene has been utilized for bio-sensing application. Huang et al. have utilized large-sized chemical vapour deposited (CVD) graphene films configured in field-effect transistors for real-time biomolecular sensing [19]. Akhavan et al. have observed the label-free detection of Single Nucleotide Polymorphisms (SNPs) of oligonucleotides with a specific sequence at a concentration of up to ~10 DNA/mL using graphene based nanostructure modified electrodes [20]. He et. al. have developed a point-of-care sensor for the detection of folic acid protein (FAP) using graphene-based SPR chips. They exploited the exceptional properties of graphene to construct a highly sensitive and selective SPR chip for folate biomarker sensing in serum. The interaction of π-stacking on the graphene coated SPR chip and the FAP analyte in serum allowed femtomolar (fM) detection of albumin mixtures [21]. Graphene has been recognised as a revolutionary material employed for sensing due to its extraordinary optical and electronic properties. These remarkable properties are derived from its unique electronic and lattice structure. Its 2D hexagonal packed structure of sp2 hybridized carbon atoms has unique band structure.

The band structure of graphene also depends on the stacking and the number of layers. The single layer graphene structure has linear bands with zero band gap. When stacked as a bilayer, the band structure changes as the linear bands fold themselves into parabolic bands. Bilayer graphene are generally classified based on stacking such as AA and AB stacking. The band structure of both the stacking are also different. The AA stacking band structure are two similar copies of monolayer graphene whereas in AB type, it has a pair of intersecting parabolic bands with additional non-intersecting parabolic bands with some vertical offset [22–24]. The delocalized π-electrons of a graphene sheet can enhance the binding with bio-analytes through π-π interaction [25]. The graphene layer over the gold surface act as a dielectric layer in the range of EM wave used to observe the SPR phenomenon. Although, it can enhance the adsorption capability of the analytes, however sensing characteristics in the aqueous medium during bio-sensing e.g. figure of (FOM) reduces [26]. Furthermore, the weak physical force of binding between metal and graphene can increase the sensing instability [27]. Therefore, if plasmonic fields can be generated in graphene layer system itself, it can serve as a plasmonic material for the generation of SPP waves and additionally it can offer enhanced bio-molecular interaction through π-π interaction leading to greater sensitivity. Inspired by this, some groups have reported SPR based sensors employing graphene as the plasmonic material [28, 29]. Maleki et al. [28] have proposed a gas sensor based on double layer graphene nanograting fabricated over a dielectric substrate. They were able to observe the device sensitivity of 430 nm/RIU over incident radiation of 1μm – 2μm. Similarly, Wu et al. [29] have proposed a SPR sensor employing graphene nanoribbons array over $SiO_2$ substrate working in the infrared regime. They observed sensitivity of 4720 nm/RIU with FOM of 5.43. They observed that increasing the stacks of graphene nanoribbons can decrease the sensitivity of the device [29]. Therefore, mere increasing the no of graphene layers would not enhance the performance of graphene-based biosensors. The band structure of graphene plays an important role here. In case of such 2D materials, increasing the number of layers alters their electronic band structure. In 2D material such as $MoS_2$, its monolayer structure has a larger band gap when compared to its bilayer system. The bilayer $MoS_2$ has a smaller indirect band gap due to strong interlayer coupling which reduces its photoluminescence property as compared to monolayer $MoS_2$ [30]. Similarly, in metal dichalcogenides ($PtTe_2$ and $PtSe_2$), Li et al. [31] have observed band gaps of 1.8 and 0.6 eV for mono- and bilayer of $PtSe_2$, respectively, and 0.5 eV for monolayer of $PtTe_2$. They also observed that increasing the number of layers can change the conductivity. Therefore, increasing the number of layers in these novel 2D materials can alter their physicochemical properties.

In the past few years, the twisted bilayer graphene (T-BLG) [32, 33] has enhanced the employability of graphene nanostructured films for various applications [34]. The T-BLG is a bilayer graphene structure with a relative in-plane twist between the two layers (Fig 2(d)). When compared to the bilayer graphene (BLG) structure, it has a very complex band structure [35]. The properties of BLG are stacking dependent (AA or AB). Interestingly, the band structure and hence the properties of T-BLG are dependent on the relative in-plane twist angle between each graphene layer of the stacking [36–38]. The twisted graphene could enhance bio-sensing applications [39]. In our previous work, it has been observed that for different relative in-plane twists, the optical properties of the T-BLG systems also changes [40]. For different incident energy photons, it behaves as either semiconducting, metallic, semi-metallic or a dielectric depending on the relative in-plane twisting angle. Therefore, it can be foreseen another important parameter (viz. relative in-plane twist angle in T-BLG) for development of next generation graphene-based sensors. The twisting of graphene layer in T-BLG system can enhance the plasmonic field at some specific angle which can be harnessed for the development of efficient bio-sensor. Here, we demonstrate that MLG, BLG and T-BLG can be employed for the generation of SPR for enhanced bio-sensing applications. It is noteworthy that this study shows sensing metrics can be controlled by altering the relative in-plane twist angle in T-BLG system. In this work, we present a theoretical simulation on a SPR biosensor employing monolayer graphene (MLG), BLG (both AA and AB stacking) and T-BLG as plasmonic materials and observe their sensing capabilities. The MLG, BLG and T-BLG can be deposited through several techniques including chemical vapor deposition (CVD), epitaxial techniques [41–43]. In our simulation, the Kretschmann configuration for wavelength interrogation mode was developed by incorporating the semi-conductor Zinc Selenide (ZnSe) as the coupling medium and MLG, BLG, T-BLG as the plasmonic materials. The SPP waves were generated using EM wave in the IR regime. We observed excellent resonance curves with a very high sensitivity and FOM. The model was tested for bio-sensing application by changing the RI of the aqueous medium. The sensitivity of the T-BLG sensor was found to be 29111 nm/RIU which is highest as compared to others reported in literature for 1° twist angle.

## 2. Simulation setup

A finite difference time domain (FDTD) method was employed for the simulation of SPR phenomenon using a commercial package of Lumerical [44, 45]. The FDTD method is highly reliable and advantageous over other techniques in solving numerically the discretized form of Maxwell's equations for complex geometries of materials. The equations can be solved numerically and simultaneously in both space and time domains. It also has the benefit of calculating reflectance for multiple wavelengths of light per simulation [46]. The FDTD can be used to calculate most of the useful quantities related to electromagnetics such as Poynting vector, electromagnetic fields of transmitted/reflected waves. The solutions in frequency domain can be obtained using the standard Fourier transform. The main advantage of FDTD compared to the other methods such as Transfer Matrix Method (TMM) and other methods is that it can give the transmission properties over a wide spectral range with just a single calculation. It can also calculate time domain field profile and the currents over the entire computational domain. Moreover, it can treat defects with no additional computational complications [47].

The FDTD simulation setup, the schematics of graphene system such as MLG, BLG and T-BLG systems and the graphene-SPR sensor setup are shown in Fig. 1-3, respectively.

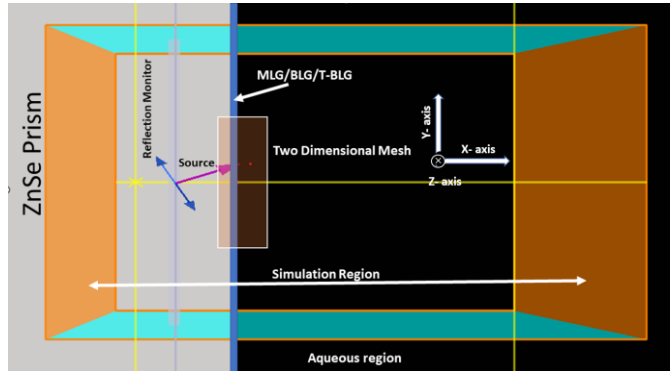

Fig.1. FDTD simulation setup depicting the various simulation parameters.

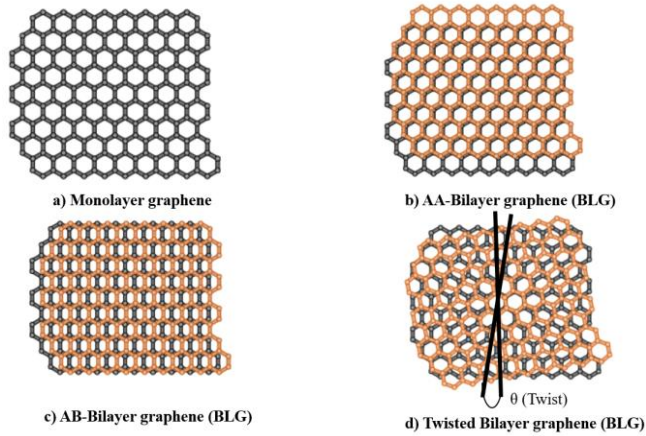

Fig. 2. The schematic representation of different graphene structures used in the simulation.

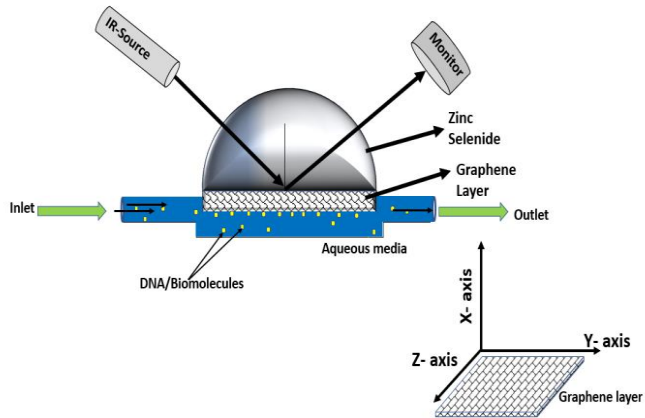

Fig. 3. Schematic for proposed Graphene-SPR sensor consists of ZnSe as coupling prism, electromagnetic wave IR source, detector as monitor. The plasmonic field is generated in graphene layer. XY is the plane of incidence and YZ is surface plane of graphene layer. The analytes are dispersed through the flow cell.

The simulation was carried out in an aqueous medium of refractive index 1.33 RIU and pH = 6. This is one of the physiological conditions where pH =6 is suitable for biosensing [14]. However, change in pH does not have direct impact on SPR measurement until it changes the RI of the aqueous medium. A plane wave source was allowed to incident at zinc selenide (ZnSe) – graphene system interface at an angle of 25° which is slightly more than the critical angle of incidence for total internal reflection. Few of the simulation parameters for the material used are mentioned in the Table 1.

| Material | Thickness | Refractive Index | Reference |
|---|---|---|---|
| ZnSe | 100 μm | 2.46 | [48, 49] |
| MLG | 0.34 nm | As per our previous work | [32] |
| BLG (AA/AB stacking) | 0.68 nm | | |
| T-BLG (for different twists) | 0.68 nm | | |
| Aqueous medium | 100 μm | 1.33 | - |

Table-1: The thickness and RI of the layers used in the simulation

The thickness of ZnSe and aqueous medium is taken very large to represent continuous medium. Although a smaller simulation region is considered to solve the Maxwell's equation at the ZnSe/Graphene interface. The perfectly matched layer (PML) boundary condition with steep angle profile of 12 layers was used to minimize refection from the boundary as the wave enters the layer. Linear discrete Fourier transform (DFT) monitors were used to capture reflected and transmitted electric field at 350 nm away from the interface. These monitor returns the Fourier transform of time-domain electromagnetic fields. To obtain the resonance condition, wavelength interrogation sweep was generated from 1μm-15μm with 1000 iterations. Later after the observation of SPR wavelength at 13.7 μm in MLG system, the sweep was reduced to 12μm – 15μm with 5000 iterations for better numerical accuracy. A mesh override was selected in the propagation direction of the plane wave to get more precise result. To observe the two-dimensional plasmonic field generated at the graphene surface, a 2D-DFT monitor was placed and plasmonic field was captured at the resonance condition. The refractive index (RI) of MLG, BLG and T-BLG was adapted as per the DFT calculation from the previous work of our group [40]. For biomolecules/analytes sensing, the refractive index of the aqueous medium was varied from 1.33 – 1.331 in steps of $10^{-4}$ RIU. The RI range was chosen to represent biosensing applications in aqueous medium. Most of the biosensors need aqueous medium where the change in RI may lie in this range [50]. The calibration curves showing the variation of shift in resonance wavelength (Δλ) as a function of change in RI of aqueous medium possessing the analytes (Δn) were obtained for MLG, BLG and T-BLG (with different twist angles) systems. Mathematically, sensitivity ($S_\lambda$) can be calculated from the slope of the calibration curve using the formula [14].

$$S_\lambda = \frac{\partial \lambda_{RW}}{\partial n} \quad (1)$$

The figure of merit can be calculated using the expression [15]:

$$FOM = \frac{S_\lambda}{FWHM} \Delta T \qquad (2)$$

where FWHM is full width half minima and $\Delta T = 1-T_{min}$, $T_{min}$ is the dip in the reflectivity at the resonance wavelength (RW).

## 3. Results and Discussion

### 3.1 Monolayer Graphene

In Kretschmann configuration with wavelength interrogation, p-polarized electromagnetic (EM) waves were allowed to incident at the ZnSe-MLG (Z/M) interface as shown in Fig. 1. For bio-sensing application, the dielectric medium is generally aqueous (RI = 1.33 RIU and pH = 6). Thus, SPR sensing module consisted of ZnSe/MLG/aqueous (Z/M/A) system. The SPR spectrum and the 2D plasmonic field distribution from the Z/M/A system with pure water medium is shown in Fig. 4.

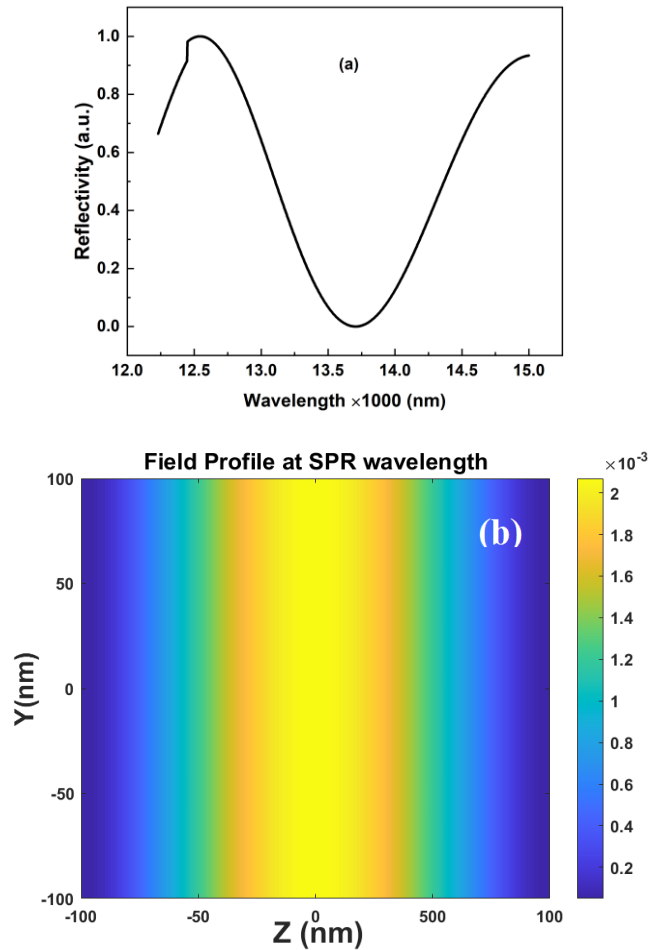

Fig. 4. (a) SPR spectrum from Z/M/A system. (b) The plasmonic field profile for the Z/M/A system obtained at resonance wavelength (on Y-Z plane).

The spectrum (Fig-4a) shows a good trend with a resonance wavelength (RW) of about 13.7μm. The spectrum is similar to that of a glass/gold/aqueous system [51]. The field distribution of 2D plasmonic field (Fig-4b) is also found to be similar to that of a glass/gold/dielectric [44]. The concentration of the bio-analytes in the aqueous medium is changed by changing the RI of the aqueous medium from 1.33 to 1.331 in the step of $1\times10^{-4}$ RIU. The corresponding SPR spectra were simulated and are shown in Fig. 5(a).

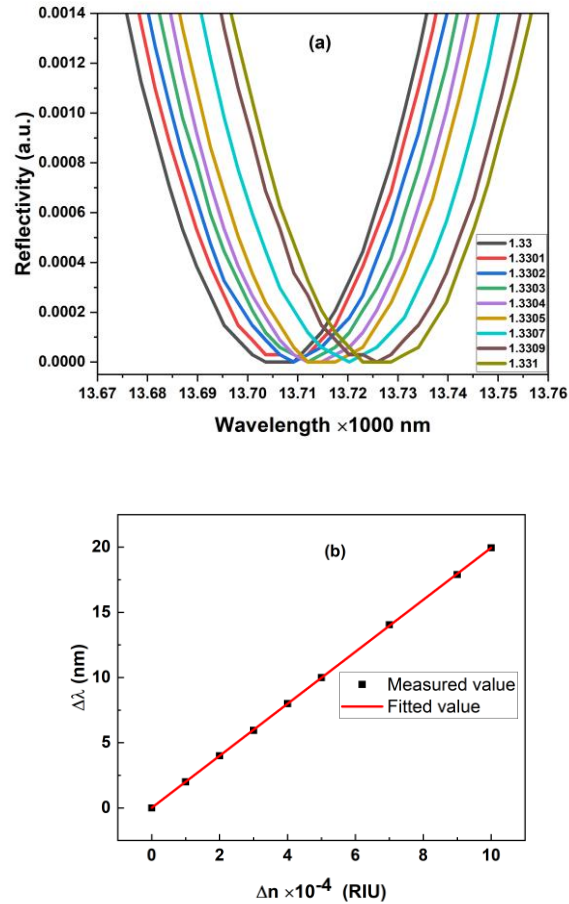

Fig. 5. (a). SPR spectra for different refractive indices of aqueous medium due to adding of biomolecules from Z/M/A system (b) Calibration curve showing change in resonance wavelength (Δλ) Vs change in RI (Δn).

The resonance wavelength (RW) shifts towards higher wavelength as a function of concentration of the bio-analytes in the aqueous medium. A calibration curve is obtained by plotting the shift in RW (Δλ) due to change in RI (Δn) of the aqueous medium. A linear trend in the calibration curve was obtained which is one of the indicators of good sensor.
The sensitivity of the device was calculated from the calibration curve (Fig. 5(b)) using the Eqn. 1, and it was found to be approximately 20,000 nm/RIU. This is about four times the average sensitivity reported yet for graphene based SPR sensors in infrared regime [28, 29]. This enhancement is due to the choice of suitable coupling medium i.e. ZnSe in the present case. The ZnSe medium provides least attenuation of the incident EM wave in IR regime which otherwise gets absorbed largely in a glass coupling medium. Thus, the MLG system over the ZnSe coupling medium can be a good candidate for sensing application using the SPR phenomenon.

## 3.2 Bilayer and Twisted Bilayer Graphene System (BLG & T-BLG)

The bilayer graphene (BLG) is considered as another category of materials which are being employed for various novel applications. Due to the structural difference between MLG and BLG, bilayer system has larger band gap and altered optical response in infrared region. The BLG structure naturally tends to have Bernal (AB) type stacking (e.g. highly oriented pyrolytic graphite). Although, using certain controlled deposition techniques such as CVD [52, 53], non-Bernal stacking (AA or twisted) can be achieved [52]. The optical properties of BLG with different types of stacking are known to be different. It is noteworthy that the coupling between the two layers of a BLG system can be altered by applying a relative in-plane twisting [54]. It is, therefore, interesting to study the effect of twisting on sensing performance of a BLG-based SPR sensor. On investigating, we found that the BLG system with AA and AB (non-twisted) stacking does not show any significant change in the RW, and sensitivity as compared to MLG system. The sensitivity of AA and AB stacked BLG are found to be 20,016 and 19300 nm/RIU which are closer to MLG system (~20000 nm/RIU). Additionally, we found that twisting of AA stacked BLG does not show any further improvement in sensitivity and FOM. This was expected from our previous studies wherein the optical properties of AA stacked BLG does not reveal significant change as compared to the untwisted state [40]. However, it was also observed that the optical properties of AB stacked BLG system was highly dependent on the in-plane twist angle. A schematic phase diagram dependent on the twist angle was proposed [40]. Therefore, it is interesting to study the SPR based sensing of the AB stacked BLG system as a function of in-plane twisting.

The T-BLG system is rather an unconventional and complex system. It possesses a complex band structure which varies for different in-plane twisting. Due to the twists, hexagonal Moirè patterns are generated with irregular defects along the bilayer system which affects its band structure leading to a twist angle dependent behaviour [40]. The AB type stacked BLG behaves like a semi-metallic in the mid infra-red region due to small angular twists (1°-10°) and as metal for larger twists like 20° and 80°. It was found that coupling between the layers for lower twist angles is high and it diminishes for the large twist angles [54]. Therefore, we performed the simulation to observe the SPR curve from AB stacked BLG system in the aqueous medium for the twist angles (Z/T/A) in two different regions viz. low angular twist (1°, 2°, 3° and 7°, semi-metallic behaviour) and high angular twist (20° and 80°, metallic behaviour). Fig 6(a) shows the SPR curves for the T-BLG systems with different twist angles. It can be observed from the figure that the spectrum shifts towards left (blue shift) of the AB stacked BLG with 0° twist. The blue-shift of RW ($-\Delta\lambda$) SPR spectra towards for the given aqueous medium suggests increase in metallicity of the T-BLG system due to twisting. The amount of blue shift in RW is calculated and shown in Fig. 6(b).

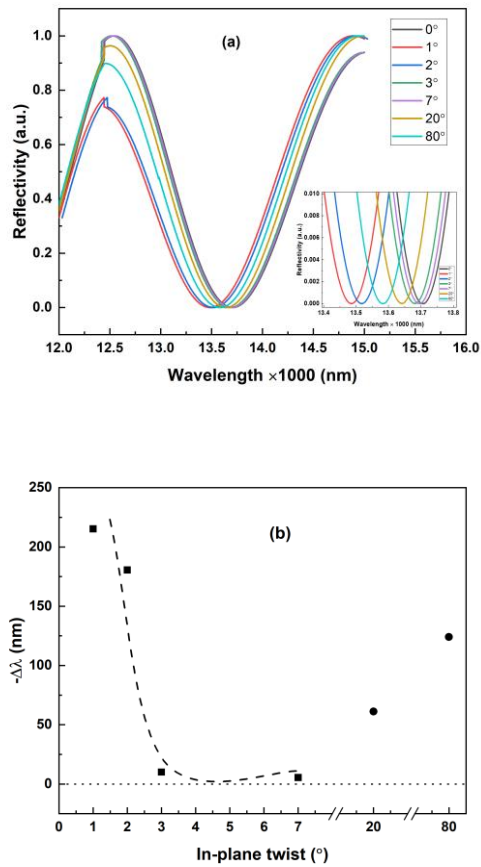

Fig. 6. (a) SPR spectra and (b) corresponding RW (Δλ) for different in-plane twist angles in AB stacking of T-BLG (Z/T/A) system. (-) sign represents blue-shift in RW of the SPR spectra as observed in (a). The dash line is drawn to show the reference. Z/T/A corresponds to ZnSe/ Twisted-BLG/ Aqueous.

It can be observed from the Fig. 6(b) that maximum blue-shift in RW (220nm) was obtained for a twist angle 1°. In the low angle regime, it decreases sharply till 3° and then saturates at 7°. The -Δλ value increases moderately for large twist angle e.g., 20 and 80°. This variation thus suggested that the T-BLG system exhibited the largest metallic characteristics for the twist angle of 1°.

The sensitivity towards bio-analytes in aqueous medium was calculated from the calibration curves obtained for the BLG and T-BLG as a function of twist angle. The results are summarized as a bar diagram in Fig. 7. The sensitivity for MLG, and AA, AB stacked BLG system with 0° twist are comparably similar. Interestingly, on twisting the layers in Z/T/A system by 1°, the sensitivity of the sensor enhanced remarkably to 29120 nm/RIU which is more than 140% as compared to MLG or BLG (with 0° twist). The sensitivity is lowest for 80° twist of the T-BLG system.

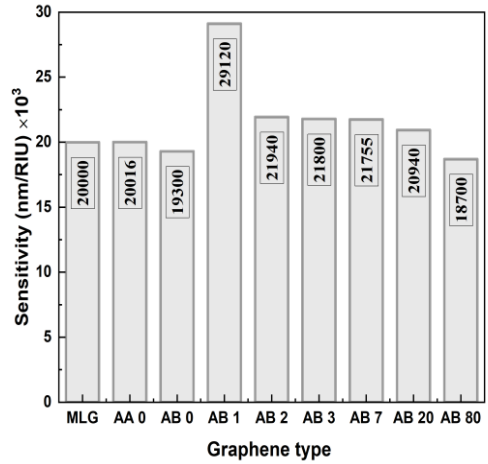

Fig. 7. Sensitivity for different graphene systems. Corresponding values of sensitivity are as mentioned

The 2D plasmonic field over the surface (Y-Z plane) of BLG and T-BLG systems was estimated from the simulation. Interestingly, the magnitude is enhanced due to the addition of extra layer BLG as compared to MLG (Fig 2(b)). The relative field strength (RFS) was calculated from the ratio of fields of a BLG system to MLG system.

$$Relative\ field\ Strength\ (RFS) = \frac{maximum\ plasmonic\ field\ strength\ of\ BLG\ or\ TBLG}{maximum\ plasmonic\ field\ strength\ of\ MLG}$$

As observed, the RFS increases to nearly 300 times on simply going from MLG to BLG system (Fig. 8). Under the influence of twist in T-BLG, the RFS is found to be maximum (~ 425) for a twist angle of 1°. It reduces to nearly to 315 on further increase of twist angle till 7°. There is a drastic decrease in RFS on higher twist angle (i.e., at 20° and 80°). This observation is in consistence with the sensitivity data as shown in Fig. 7.

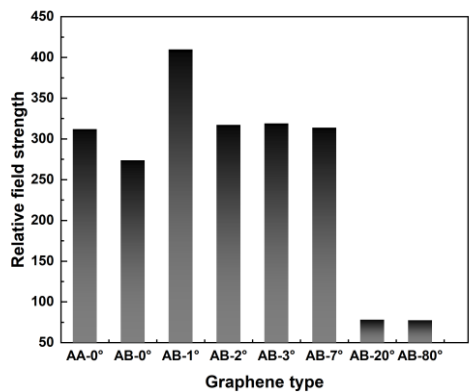

Fig. 8. Relative plasmonic field strength (RFS) of BLG and T-BLG system in comparison to MLG system

The FOM was calculated for the BLG and T-BLG systems with different values of relative in-plane twist angles using Eqn. 2. This is shown in Fig. 9.

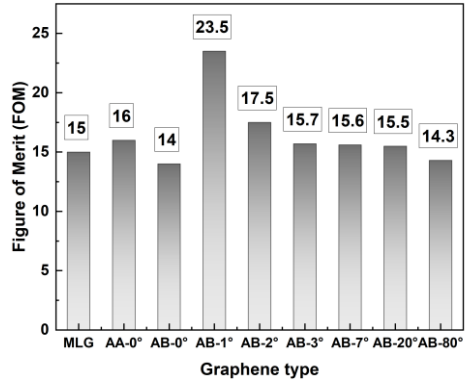

Fig. 9. Figure-of-merit (FOM) for different graphene systems.

The largest and the smallest values of FOM was obtained for 1º and 80º twist angle in the T-BLG system respectively. This is in consistent with the other parameters.

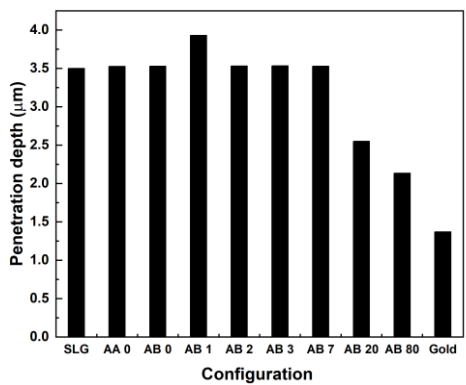

Fig. 10. Penetration depth of evanescent waves for different systems.

The penetration depth was measured for all the graphene layer systems and then compared with the classical BK-7/Gold based Krescthmann configuration SPR design (Fig. 10). The distances were computed from the interface up to 10% of the maximum amplitude of the evanescent waves. In consistent with the previous results, 1° twisted BLG system has the greatest penetration depth when compared to the other systems. The proposed graphene-SPR sensor has a greater penetration depth than the classical BK7/Gold.

The T-BLG system possessing twist angle closer to 1° has been studied widely for its extraordinary behaviour. This angle of twist is also called as magic angle. At this magic angle the T-BLG system shows superconducting behaviour [55] as it exhibits ultra-flat bands near charge neutrality. For the twist angles close to the magic angle such as 1°, the energy separation between the nearly flat bands and the nearest conduction and valence bands is ~100 meV which is comparable to energy of mid infrared regime (80 meV-200 meV). These energies are comparatively larger than the energy separation between the pair of nearly flat bands at the magic angle [56]. When light impacts on T-BLG, its time-periodic electric field vibrates electrons around their equilibrium positions and induces an interband transition. When a plane polarized EM wave is incident, the electrons will build up an oscillating charge density with the same wavevector. This oscillating charge density in turn, creates an oscillating electric field that develops an external field, and at resonance this induced field is strong enough to maintain the oscillation for a longer duration. This resonance behaviour gives rise to collective modes that are called interband plasmons [57]. Thus, due to this strong collective mode oscillation, the interaction with the nearby dielectric is enhanced [58]. Therefore, the π-π interaction with the analytes present in the sensing medium will also be high for the T-BLG near magic angle. Now as the twist angle moves further away the magic angle, the oscillation of this collective mode might decrease resulting in weaker interaction. Also, a strong coupling between the two layers of a BLG system is reported at the magic angle which reduces on increasing twist angle. In the present case, the decoupling is found near to 7° twist angle (Fig. 6(b)). For the very high twist angle (20° and 80°), the coupling increases to about 30-50% of the maximum. This is due to the traditional metallic behaviour of T-BLG system as reported in [40].

Table-2 Comparison of the proposed work with the preliminary literature available

| Glass | Metal | Sensing layer | Sensitivity | Reference |
|---|---|---|---|---|
| BK7 | Au | - | 2161nm/RIU | [59] |
| Optical fiber | Au | - | 2280nm/RIU | [15] |
| Optical fiber | Ag | - | 3710nm/RIU | [15] |
| Optical fiber | Ag | Graphene (10 layers) | 5000nm/RIU | [15] |
| Fluorinated $SiO_2$ | - | Bilayer graphene nanograting | 430nm/RIU | [28] |
| Quartz | - | Graphene ribbon | 4720nm/RIU | [29] |
| | | array | | |
| ZnSe | - | Monolayer graphene | 20,000 nm/RIU | This work |
| ZnSe | - | Twisted Bilayer Graphene | 29,120 nm/RIU | This Work |

## Conclusion

SPR sensors employing graphene as functional layer can be considered as next generation device wherein the performance can be extended to a new higher limit. The graphene layer in SPR sensor can be utilized for dual purpose like a plasmonic material for generation of SPP waves and functional layer for the enhancing adsorption of analytes through π-π interaction. In this article, a Kretschmann based SPR phenomenon in MLG, BLG and T-BLG layers was simulated and their sensing performance towards bio-analytes in aqueous medium was compared. The MLG system can be used for bio-sensing application however, the sensing merits can be pushed to next level by using T-BLG system with a provision to induce relative in-plane twist between the layers of AB stacked BLG. The sensing performance of T-BLG (AB stack) with 1º twist angle (~ magic angle) in terms of sensitivity, RFS and FOM was found to be superior as compared to other states of either MLG or BLG. The bilayer graphene system in the vicinity of magic angle can exhibit strong interlayer coupling leading to exceptionally flat bands which can support highly correlated phenomena. This may lead to the enhance performance of T-BLG (with 1º twist angle) in a SPR based sensing device. A comparison of the proposed work with the available literature is shown in Table 2.

## Acknowledgements

Author A.K would also like to acknowledge DST-INDIA, SERB project (CRG/2018/000755) for fellowship and BITS PILANI, for the Infrastructure and Department of Physics for funding for Lumerical software. This is a post-peer-review, pre-copyedit version of an article published in Plasmonics. The final authenticated version is available online at: https://doi.org/10.1007/s11468-022-01760-2

## Declarations

**Ethical approval**
Not applicable

**Authors' contribution**

Simulation, Data analysis and manuscript preparation were done by Amrit Kumar. Conceptualization, data analysis and manuscript preparation were done by Raj Kumar Gupta. Data analysis and manuscript preparation were done by Manjuladevi V.

**Competing interests**

There is nothing to be declared.

**Data availability**

Data to be made available on demand to the corresponding author.

**Funding**
Not applicable

**Code availability**
Not applicable